\documentclass[a4paper,12pt]{article}
\input{epsf}
\usepackage{latexsym,indentfirst}
\begin{document}
\newcommand{\orh}{{\overline \rho}}
\newcommand{\oet}{{\overline \eta}}
\newcommand{\be}{\begin{equation}}
\newcommand{\ee}{\end{equation}}
\newcommand{\bea}{\begin{eqnarray}}
\newcommand{\eea}{\end{eqnarray}}
\pagestyle{plain}
\title{Is the quark- mixing matrix moduli symmetric?}
\author{ S. Chaturvedi \thanks{scsp@uohyd.ernet.in}\\
\rm School of Physics, University of Hyderabad, \\
 Hyderabad 500 046 India\\
Virendra Gupta \thanks{virendra@aruna.mda.cinvestav.mx}\\
\rm Departamento de Fis\'ica Aplicada, CINVESTAV-Unidad M\'erida\\
 A.P. 73 Cordemex 97310 M\'erida, Yucatan, Mexico}
\date{\today}
\vskip-1cm
\maketitle
\begin{abstract}
If the unitary quark- mixing matrix, $V$, is moduli symmetric
then it depends on three real parameters. This means that there
is a relation between the four parameters needed to parametrize
a general $V$. It is shown that there exists a very simple relation 
involving $|V_{11}|^2,~|V_{33}|^2,\orh
$ and $\oet$. This relation is compared with the present experimental data.
It is concluded that a moduli symmetric $V$ is not ruled out.
\end{abstract}
\newpage
\section{Introduction}
It is well known that for three generations, the general parametrization
\cite{1},\cite{2} of the Cabibbo-Kobayashi-Maskawa (CKM) quark-mixing
matrix, $V$, depends on four parameters, namely, three angles and a
phase. Experimental data gives the values of the moduli $|V_{ij}|$ and a
particular parametrization of $V$ is needed to determine the complex
matrix elements of the unitary matrix $V$. Four moduli (obtained from data)
are needed to determine the four parameters in $V$.

The present data \cite{2}, gives us the ranges of $|V_{ij}|$. These are
    \bea
    V_{EXP}=\left(
    \begin{array}{ccc}
    0.9741-0.9756&0.219-0.226&0.0025-0.0048\\
    0.219-0.226&0.9732-0.9748&0.038-0.044\\
    0.004-0.014&0.037-0.044&0.9990-0.9993
    \end{array}\right).
   \label{1}
   \eea
It is clear from these values that there is a possibility that $V$ might
turn out to be moduli symmetric. The ranges suggest that $|V_{ij}|
=|V_{ji}|$ for $(i,j)=(1,2)$ and $(2,3)$, but it seems that $|V_{13}|
\neq |V_{31}|$. However, the latter matrix elements are difficult to
measure and may change in future. Since $V$ is unitary, it follows that
\be
\Delta \equiv |V_{12}|^2-|V_{21}|^2=|V_{23}|^2-|V_{32}|^2|=V_{31}|^2-
|V_{13}|^2.
\label{2}
\ee
So either $V$ is completely moduli symmetric  (~$\Delta=0$~) or it is fully
asymmetric ($\Delta ~\neq ~0$).  Recently an attempt to understand the
smallness of this asymmetry (i.e. smallness of $\Delta$) has been made
\cite{3}.

In this note, we explore the experimental consequences of a moduli
symmetric $V$, denoted by $V_{MS}$. Since, $\Delta=0$ for $V_{MS}$,
this gives an extra condition\footnote{An explicit parametrization for 
$V_{MS}$ was considered inreferences 5 and 6. A relation involving 
$|V_{12}|, |V_{23}|$ and theparameters $\orh$ and $\oet$ of the unitarity 
triangle was pointed out.}and consequently a general parametrization
of $V_{MS}$ contains only three real parameters \cite{3},~\cite{4}.

The important point is that if $V=V_{MS}$ then there will be a relation
between four measurables, which for a general $V$ would be independent.
In this note we obtain a general relation and confront it with available
data. 
 \section{The relation}
There is a lot of interest in measuring the quantities connected with
the unitarity relation or triangle, viz., 
\bea
       &&V_{11}V^*_{13} + V_{21}V^*_{23} +V_{31}V^*_{33} =0,
\label{3}
\eea
Define $z_{i}$=$V_{i1}V^*_{i3}~; i=1,~2,~3$  then Eq.$(\ref{3})$ can be
written as
\be
            -z_{1}/z_{2} - z_{3}/ z_{2}=1.
\label{4}
\ee
 This defines a triangle. Define the complex numbers\cite{2}
\be
-z_{1}/z_{2}= \orh +i\oet,
\label{5}
\ee
so using Eq.$(\ref{4})$,
\be
-z_{3}/z_{2}=(1-\orh)-i\oet
\label{6}
\ee
This notation like that for the angles of the triangle has become standard. 
The angles
$\alpha=arg(-z_{3}/ z_{1})$, $\beta=arg(- z_{2}/z_{3})$, and
$\gamma=arg(-z_{1}/z_{2})$  of the triangle satisfy
\be
  \sin\alpha=\frac{\sin\beta}{\sqrt{\orh^2+\oet^2}}=
   \frac{\sin\gamma}{\sqrt{(1-\orh)^2+\oet^2}},
\label{7}
\ee
and
\be
         \tan\gamma = \oet/\orh .
\label{8}
\ee
To obtain the desired relation we note that from Eqs.$(\ref{5},\ref{6})$
\be
\frac{(1-\orh)^2+\oet^2}{\orh^2+\oet^2} =
\frac{|V_{33}V_{31}|^2}{|V_{11}V_{13}|^2} =
\frac{|V_{33}|^2}{|V_{11}|^2}\equiv r
\label{9}
\ee
The last equality follows if $V$ is moduli symmetric since then
$|V_{13}|^2= |V_{31}|^2 $. Thus, for $V_{MS}$, the four independent
quantities $\orh,\oet,|V_{11}|$ and $|V_{33}|$ are related.

To compute the ratio $r$, we convert the ranges for $|V_{ij}|$ given in
Eq.$(\ref{1})$ into a central value with errors. This procedure gives,
$|V_{11}|=0.97485\pm 0.00075$, $|V_{33}|=0.99915\pm 0.00015$
$|V_{13}|=0.00365\pm 0.00115$ and $|V_{31}|=0.009\pm 0.005$. Using these we
find for $V_{MS}$, $r_{MS}= |V_{33}|^2/|V_{11}|^2= 1.05048\pm 0.00165$, 
otherwise
$r= |V_{31}V_{33}|^2/|V_{13}V_{11}|^2= 6.38683\pm 8.15826$. The extremely 
large error in $r$ reflects the large errors in $|V_{13}|$ and $|V_{31}|$.

According to the relation in Eq.$(\ref{9})$, $\orh$ and $ \oet$ 
lie on the circle
\be
(\orh+1/(r-1))^2+\oet^2=(\sqrt{r}/(r-1))^2
\label{10}
\ee
The circles for $r_{MS}$ are plotted in Fig 1. The relevant portion
in the first quadrant is shown  since $\orh$ and $\oet$ are both positive.
It should be noted for $r=1$ the circle degenerates into a straight line 
$\orh=1/2$. As $r$ increases, the radius increases and the centre approaches 
the origin along negative $\orh$-axis. For $r=6.38683$ the centre of the 
circle is at $(-0.185638,0)$ and the radius is $0.469148$. 
Since  there is a  large error in $r$ (for the asymmetric case), 
it is clear that the range of values for $r$ contain those for $r_{MS}$. 
Given the present data it seems that the possibility that $V$ is moduli
symmetric is not ruled out. Our point here is that Eq.$(\ref{10})$ provides
a very simple and direct way to check if $V$ is moduli symmetric or not. 
One has to await more accurate data for $|V_{13}|$ and $|V_{31}|$ to come to a 
definitive conclusion in this regard. 

From Eqs,$(\ref{7}, \ref{8})$, we can determine
\be
   \sin 2\beta = \frac{2\oet(1-\orh)}{(1-\orh)^2+\oet^2}.
\label{11}
\ee
The curve in Eq.$(\ref{11})$ represents the product of straight lines given by 
\bea 
\oet&=&\tan\beta (1-\orh),\label{12}\\
 \oet&=&\cot\beta (1-\orh).\label{13}
\eea
Experimentally, different groups and different decay modes give a wide
range of values for $\sin 2\beta$. Using the average of all modes and
groups \cite{7}, $\sin 2\beta=0.699\pm 0.054$, the straight lines  in
Eq.$(\ref{12}, \ref{13})$ are also plotted in Fig.1. It is interesting to 
note that the circles for  $r=r_{MS}=1.05048\pm 0.00165$ and the line 
in Eq(13) with $ \cot \beta= 2.45368 \pm 0.265066$  have  
a small region of intersection around 
$\orh=0.447577$, $\oet=1.36391$. However, this is excluded by constraints from 
other data \cite{2}. For the general case, taking 1/2 the error into account, 
that is $r=6.38683\pm 4.07913$, one finds that there is a 
large region of intersection region with the lines in Eq.$(\ref{12}$ with
$\tan \beta = 0.407551\pm 0.044027$, though there is no intersection with  
Eq.$(\ref{13})$. As one can see the lines corresponding to Eq.$(\ref{12})$ 
with $\tan \beta = 0.407551\pm 0.044027$ have a small region of 
intersection with the circles for 
$r=r_{MS}=1.05048\pm 0.00165$ around the point $\orh=0.492779$  and 
$\oet=0.206728$ keeping open the possibility that $V$ be moduli symmetric.

Further, we note that $\sin^2\gamma/\sin^2\beta=r$ so that knowledge of
$\beta$ from $\sin 2\beta$ enables one to obtain angles
$\alpha$ and $\gamma$.  The values of the angles for $r=r_{MS}$ and the 
general $r$ are given in Table I. The values in the two columns, as expected, 
are fairly different. The point to note is that for the moduli symmetric 
case, since $r_{MS}\approx1$,  one expects $\beta \approx 
\gamma$, unlike the general or
asymmetric $V$ where the two angles can be quite different (viz. Table I). It 
is very interesting to note  the value of $\sin 2\alpha$
(which is in the process of being measured) in the two cases. From Table I, 
for the central values, one expects $\sin 2\alpha =- 0.9974$ for the moduli 
symmetric case in contarast to a value of $0.163$ for a asymmetric $V$.
An experimental value of $\sin 2\alpha$  near $-1$ would favour a moduli 
symmetric $V$. 

In conclusion, we have pointed out that a simple, model independent relation 
between $\orh, \oet, |V_{11}|$ and $|V_{33}|$ provides a direct test of the 
moduli symmetry of $V$. The present data does not rule out such a symmetry. 
For a conclusive answer we must await future data. 

In our view a moduli symmetric quark-mixing matrix would be far more elegant 
and physically interesting than one with a tiny, difficult to explain, 
asymmetry.  

\vskip1.0cm  
\indent \textbf{Acknowledgements}. The  work in this paper
was done while one of us (VG) was visiting the School of Physics, University 
of Hyderabad, Hyderabad (India) during August 2003.

\newpage
\begin{tabular}{|c|c|c|}  \hline \hline
ANGLES  & $r=r_{MS}=1.05048 \pm  00165$ & $r= 6.38683 \pm (8.15826)/2$ \\
 \hline
$ \beta$ & $22.1734 \pm 2.16325^\circ$& $22.1734 \pm 2.16325^\circ$ \\ \hline
$\gamma$ & $22.7568 \pm 2.17246^\circ$ & $72.5159 \pm 58.4675^\circ$\\ \hline
$\alpha $ & $137.07 \pm 3.06581^\circ$ & $85.3107 \pm 58.5075^\circ$ \\ \hline
\end{tabular}
\vskip0.5cm
Table I  Numerical values of the unitarity triangle angles with 
errors corresponding to $r=r_{MS}$ and for general $r$. Note that in the 
latter case we have taken the error in $r$ to be half of its  actual value. 
\vskip0.5cm
\begin{figure}
\leavevmode
\epsfxsize=5.0in
\epsfbox{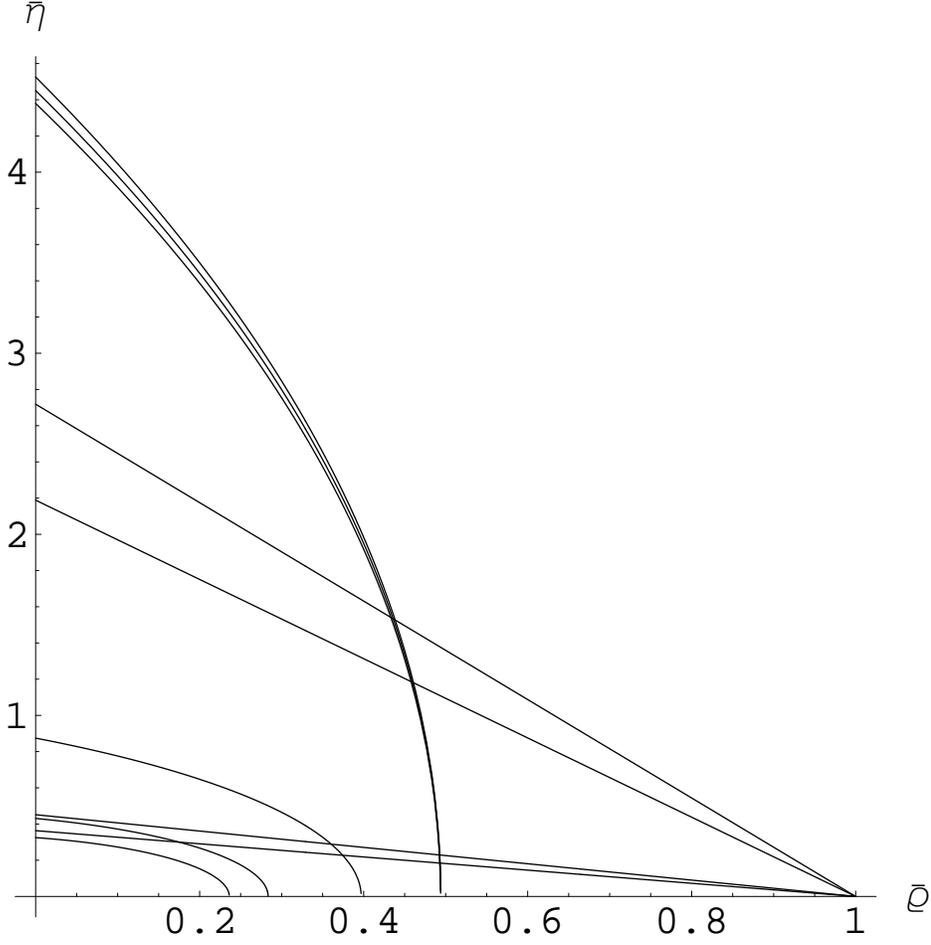}
\caption{Plots of $\oet$ versus $\orh$ : (a)  General Case: 
The lower three curves represent 
 Eq$(10)$ for $r=6.38683+(8.15826)/2, 6.38683$ and $6.38683-(8.15826)/2$. 
They are parts of circles of radii $0.341763,0.469148, 1.1617$ 
with centres at $(-0.105643,0),( -0.185638,0), (-0.764701,0)$ respectively. 
(b)  Moduli Symmetric Case: The upper  three curves again represent 
 Eq$(10)$ for $r=1.05048+0.00165,1.05048 $ and $1.05048-0.00165$. 
The radii of these  circles are  $19.6765, 20.3037, 20.9733$ 
with centres at $(-19.1828,0), (-19.8098,0), (-20.4792,0)$ respectively.
(c) The lower pair of straight lines corresponds to Eq.$(12)$ 
with $\tan \beta=0.407551\pm 0.044027$ while  the upper pair of straight lines 
 corresponds to Eq.$(\ref{13})$ with  $\cot \beta=2.45368 \pm 0.265066$.}
\end{figure}
\end{document}